\documentclass[tradiabstract]{aa}
\usepackage{graphicx}
\usepackage{amsbsy}
\usepackage{amsmath}
\usepackage{natbib}
\usepackage{color}
\usepackage{txfonts}
\usepackage{url}
\usepackage{bm}

\DeclareMathSymbol{\varOmega}{\mathord}{letters}{"0A}
\DeclareMathSymbol{\varSigma}{\mathord}{letters}{"06}
\DeclareMathSymbol{\varPsi}{\mathord}{letters}{"09}
\DeclareMathSymbol{\varPhi}{\mathord}{letters}{"08}
\DeclareMathSymbol{\varGamma}{\mathord}{letters}{"00}

\newcommand{\maps}[1]{Meteoritics \& Planetary Science}

\begin{document}

\title{Anatomy of rocky planets formed by rapid pebble accretion \\ I.\ How icy
pebbles determine the core fraction and FeO contents}
\titlerunning{Anatomy of rocky planets formed by rapid pebble accretion I}

\author{Anders Johansen\inst{1,2}, Thomas Ronnet\inst{2}, Martin
Schiller\inst{1}, Zhengbin Deng\inst{1} \& Martin Bizzarro\inst{1}}
\authorrunning{Johansen et al.}

\institute{$^1$ Center for Star and Planet Formation, GLOBE Institute,
University of Copenhagen, \O ster Voldgade 5-7, 1350 Copenhagen, Denmark \\ $^2$ Lund Observatory, Department of Astronomy and Theoretical
Physics, Lund University, Box 43, 221 00 Lund, Sweden, \\e-mail:
\url{Anders.Johansen@sund.ku.dk}}

\date{}

\abstract{We present a series of papers dedicated to modelling the accretion and
differentiation of rocky planets that form by pebble accretion within the
lifetime of the protoplanetary disc. In this first paper, we focus on how the
accreted ice determines the distribution of iron between the mantle (oxidized
FeO and FeO$_{1.5}$) and the core (metallic Fe and FeS). We find that an initial
primitive composition of ice-rich material leads, upon heating by the decay of
$^{26}$Al, to extensive water flow and the formation of clay minerals inside
planetesimals. Metallic iron dissolves in liquid water and precipitates as
oxidized magnetite Fe$_3$O$_4$. Further heating by $^{26}$Al destabilizes the
clay at a temperature of around 900 K. The released supercritical water ejects
the entire water content from the planetesimal. Upon reaching the silicate
melting temperature of 1,700 K, planetesimals further differentiate into a core
(made mainly of iron sulfide FeS) and a mantle with a high fraction of oxidized
iron. We propose that the asteroid Vesta's significant FeO fraction in the
mantle is a testimony of its original ice content. We consider Vesta to be a
surviving member of the population of protoplanets from which Mars, Earth, and
Venus grew by pebble accretion. We show that the increase in the core mass
fraction and decrease in FeO contents with increasing planetary mass (in the
sequence Vesta -- Mars -- Earth) is naturally explained by the growth of
terrestrial planets outside of the water ice line through accretion of pebbles
containing iron that was dominantly in metallic form with an intrinsically low
oxidation degree.}


\maketitle

\section{Introduction}

How our Earth formed is a question of major importance in contemporary
astrophysics, cosmochemistry, and geochemistry, not only to explain the origin
of our own planet but also to understand the prevalence of potentially habitable
planets orbiting other stars
\citep{Anglada-Escude+etal2016,Gillon+etal2017,Raymond+etal2021}.  The classical
theory of terrestrial planet formation was developed around mutual collisions
within a population of large protoplanets and smaller planetesimals.  The
runaway phase of planetesimal accretion terminates when the protoplanets reach
the oligarchic growth phase, characterized by a low number of regularly spaced
planetary embryos orbiting within an excited planetesimal population
\citep{KokuboIda1998}. Subsequent growth occurs by orbital perturbations and
giant impacts over approximately 30--100 Myr \citep{Chambers2001}. The
moon-forming giant impact is envisioned to be the last major mass contribution
to Earth \citep{CanupAsphaug2001}.

If the terrestrial planets did form through planetesimal accretion, then it
should in principle be possible to match the composition of Earth to an extant
asteroid or meteorite class that is a remnant of the planetesimal population
that formed around 1 AU in the solar protoplanetary disc. The enstatite
chondrites, a class of very chemically-reduced meteorites associated with the
inner Solar System \citep{Warren2011}, do have the right isotopic composition of
elements such as Cr and Ti, but the moderately volatile elements (such as Na and
K) and moderately refractory elements (such as Si) in the enstatite chondrites
are much less depleted than on our Earth
\citep{Javoy1995,Javoy+etal2010,Braukmuller+etal2019}. It has therefore been
proposed that the enstatite chondrites constitute a sister clade to the unknown
reservoir that formed Earth \citep{Dauphas2017,Morbidelli+etal2020}.

In contrast, the isotopic composition of Earth can also be matched by an
approximately equal contribution of inner Solar System material, with meteorites
from the early formed and thermally processed ureilite parent body as one
extreme endmember and outer Solar System material akin to CI chondrites as the
other \citep{Schiller+etal2018,Schiller+etal2020}. This combination of inner and
outer Solar System material intriguingly also agrees with the depletion of
moderately volatile elements of Earth compared to the solar abundance: the
depletion pattern of volatile elements on Earth has a similar shape to that
found for carbonaceous chondrites
\citep{PalmeO'Neill2003,Mahan+etal2018,Braukmuller+etal2019}, with a particular
affinity for the chondrule component of these meteorites
\citep{Amsellem+etal2017}. Chondrules from carbonaceous chondrites in fact have
elevated Mg/Si ratios similar to Earth \citep{HezelPalme2010}; thus the thermal
processing that caused Earth's low Si abundance may have occurred already during
the chondrule formation step rather than being inherited from accretion of
internally heated planetesimals that experienced evaporative mass loss of both
volatile elements and moderately refractory elements \citep{Hin+etal2017}.

The pebble accretion model was developed mainly to explain the formation of the
cores of gas-giant planets within the lifetime of the gaseous protoplanetary
disc \citep{OrmelKlahr2010,LambrechtsJohansen2012}.  \cite{Johansen+etal2015},
\cite{Schiller+etal2018}, \cite{Schiller+etal2020} and \cite{Johansen+etal2021}
found evidence from cosmochemical isotope data and pebble accretion theory that
the Earth formed mainly by the accretion of chondrules and primitive
dust-aggregate pebbles. A pebble accretion model is appealing to match the
isotopic constraints since the incorporation of pebbles of outer Solar System
composition is naturally achieved by the radial drift of these pebbles
\citep{Schiller+etal2018}. Importantly, the pebble accretion model connects the
formation of terrestrial planets to the formation of super-Earths, which is a
widespread outcome of planet formation around other stars, with these two
different outcomes of planet formation only being distinguished by the magnitude
of the pebble flux through the protoplanetary disc \citep{Lambrechts+etal2019}.

The thermal processing of highly volatile species, such as H$_2$O and organics,
in the envelope of a planet growing by pebble accretion naturally limits the
accreted volatile budgets to amounts similar to what is inferred for Earth
\citep{Johansen+etal2021}. This implies that terrestrial planets could have
formed outside of the water ice line, without invoking ice-poor conditions in
this cool region \citep{Morbidelli+etal2016}. The amount of water and carbon
delivered to the growing planets by pebble snow will be relatively constant, in
contrast to the traditional picture of volatile delivery through impacts that
deliver a stochastic water amount with an intrinsically large variation
\citep{Raymond+etal2004}. In contrast to the water ice, more refractory minerals
such as enstatite and forsterite (MgSiO$_3$ and Mg$_2$SiO$_4$, the main carriers
of magnesium and silicon) and iron (either pure or bound with sulfur to form
FeS) survive down to the planetary surface without significant sublimative mass
loss, when planets are in the mass range of the terrestrial planets
\citep{Brouwers+etal2018}.

Pebble accretion also differs from classical terrestrial planet formation models
by its rapid accretion timescale, which implies that the continuous release of
accretion energy is the major contributor to the temperature distribution within
a growing protoplanet. The formation of the terrestrial planets beyond the water
ice line furthermore implies that the growing protoplanet initially contains a
significant ice fraction. \cite{Lichtenberg+etal2019} modelled the efficient
loss of water from planetesimals after heating and melting by decay of the
short-lived radionuclide $^{26}$Al. However, chemical reaction between water and
metal also leads to the formation of oxidized iron (denoted here FeO although it
covers more oxidized FeO$_{1.5}$ as well) that resides in the silicates and does
not enter the core. Hence planetary growth exterior of the water ice line will
have important consequences for the core mass fraction of terrestrial planets.
The FeO fraction of the building blocks of the terrestrial planets is normally
considered to be an increasing function of the distance from the Sun
\citep{RighterO'Brien2011,Rubie+etal2015}, inspired by the low FeO fraction of
Mercury and the enstatite chondrites representing the inner Solar System and the
high FeO fraction of Vesta and the carbonaceous chondrites representing the
outer Solar System. The core mass fraction thus becomes a decreasing function of
the distance from the Sun in the classical terrestrial planet formation model.

The pebble accretion model for terrestrial planet formation was proposed very
recently but has already been challenged, particularly on the cosmochemical
evidence for an outer Solar System material contribution to Earth. The isotopic
composition of Earth and Mars plot among meteorites from the inner Solar System
(the non-carbonaceous chondrites) but lean towards meteorites from the outer
Solar System (the carbonaceous chondrites) in the major lithophile (rock-loving)
elements such as Cr, Ti and Ca
\citep{Trinquier+etal2009,Warren2011,Schiller+etal2020}. However, other elements
tell a more complex story. For the siderophile trace element Mo, whose isotope
signature in Earth is controlled by the last few percent of accretion
\citep{Dauphas2017}, bulk silicate Earth falls between an outer and inner Solar
System isotope composition -- but such mixing seems to require an s-process rich
source that has not been identified in the meteoritic record so far
\citep{Budde+etal2016}. Two-isotope plots of Mo show a line of depletion of
s-process enriched Mo that likely resides within presolar SiC grains formed in
outflows from AGB stars \citep{Dauphas+etal2002,SandersScott2021}.

The lack of outer Solar System meteorites enriched in s-process Mo relative to
the Earth led \cite{Burkhardt+etal2021} to propose that Earth formed from an
unseen reservoir of planetesimals, thereby 'refuting' a pebble accretion
origin of Earth. The matrix of CV chondrites is nevertheless enriched in
s-process Mo while the chondrules are strongly depleted \citep{Budde+etal2016},
likely due to thermal destruction of SiC grains in the chondrule formation
process. This raises the question why the CI chondrites, otherwise considered to
represent the final accretion of Earth from primitive outer Solar System
material \citep{Schiller+etal2020}, are not similarly enriched as CV matrix in
s-process Mo. We speculate here that CI chondrites have undergone substantial
heating and aqueous alteration and that matrix therefore became mixed with
dissolved chondrules, in line with other studies showing that CI may not
represent the most primitive material \citep{Asplund+etal2021}. The CV chondrite
parent body, on the other hand, experienced less water flow and hence its matrix
maintained a primordial signature of s-process enriched Mo. Using CV matrix as
representative of outer Solar System material, Mo isotopes then actually do
record an outer Solar System contribution to Earth -- and this makes models
constructed to {\it avoid} outer Solar System contamination of Earth unnecessary
\citep{Izidoro+etal2021a,Mah+etal2021,Izidoro+etal2021b,Morbidelli+etal2021}. In
the pebble accretion model, the presence of Jupiter in the Solar System mainly
affected terrestrial planet formation by preventing large chondrules and CAI
from passing its gap \citep{Desch+etal2018,Haugbolle+etal2019} while at the same
time allowing small dust aggregates to drift to the inner Solar System where it
recoagulated to form pebbles \citep{Drazkowska+etal2019,Liu+etal2022}. These
pebbles then fed the final accretion stages of the terrestrial planets.

This paper is the first in a series of papers where we explore the interior
structure and outgassed atmosphere of rocky planets that formed by pebble
accretion.  The pebble accretion model has a number of important differences
from classical terrestrial planet formation: (i) the terrestrial planets in the
Solar System formed exterior of the water ice line, (ii) water, carbon and
nitrogen are delivered in predictable amounts through accretion of cold pebbles,
(iii) the immense pebble accretion energy leads to formation of a global magma
ocean on the growing planets, (iv) metal separates from the silicates in this
basal magma ocean to form a core and mantle during the main accretion phase and
(v) the early-delivered volatiles are continuously distributed between core,
mantle and atmosphere by their partition coefficients and solubility. This
coupling, together with early mass loss driven by XUV radiation from the young
star, defines the composition of the outgassed atmosphere and hence the
conditions for prebiotic chemistry on the planetary surface.

We start here in Paper I by describing the core design of our planetary
differentiation code ADAP (Accretion and Differentiation of Asteroids and
Planets). We use the code first to model the thermal evolution and
differentiation of a planetesimal with a radius of 250 km, to demonstrate that
Vesta is a surviving protoplanet that lost out in the competition for pebbles to
other protoplanets that grew to form Mars, Earth and Venus. We show that the
melting of Vesta's primordially accreted ice led to oxidation of the iron to
form magnetite.  Coupling this model to the terrestrial planet growth tracks of
\cite{Johansen+etal2021}, we show that the FeO mass fraction of a planet is a
decreasing function of the planetary mass and that the FeO contents of Vesta,
Mars and Earth all fit with the expectations from early oxidation by interaction
with water after the seed protoplanet melted, combined with later accretion of
relatively reduced material whose iron is predominantly (approximately 90\%) in
metallic form. The core mass fraction thus becomes an increasing function of the
planetary mass -- and this way pebble accretion naturally provides a prediction
for how the core mass fraction depends on the planetary mass, without invoking
a spatially dependent oxidation state of the source material.

This paper is organized as follows. In Section 2 we describe the basic
functionality of the ADAP code in terms of planetary heating and
differentiation. We present the results from the code in Section 2 on the
internal evolution of Vesta-sized planetesimals that form with various initial
abundances of radioactive $^{26}$Al/$^{27}$Al. We use these simulations to map
out the time-dependent formation regions of different planetesimal types
characterized by their degree of internal differentiation and their mantle FeO
contents. In Section 4 we apply the results of Section 3 to calculate the core
mass fraction and FeO mantle fraction for terrestrial planets that form by
pebble accretion using the growth tracks presented in \cite{Johansen+etal2021}.
We find that the FeO mass fraction is a decreasing function of planetary mass,
while the core mass fraction shows the opposite trend. These results are in good
agreement with the measured core mass fractions and FeO mantle fractions of
Vesta, Mars and Earth. Finally, in Section 5 we summarize the results of Paper
I.

\section{The ADAP code}

We present here the ADAP code (Accretion and Differentiation of Asteroids and
Planets). The code solves the 1-D thermal conduction equation on a spherically
symmetric grid. Importantly, the code includes the time-dependent melting of the
layers and the sinking of denser material through lighter material, while
conserving the total energy. The code considers three heat sources: (1) decay of
short-lived radionuclides, (2) accretion and (3) differentiation.

\subsection{Basic materials}

ADAP bundles, for simplicity, the interior composition of asteroids and planets
into five basic materials:
\begin{enumerate}
  \item Metal (Met)
  \item Silicate (Sil)
  \item Water (Wat)
  \item Clay (Cla)
  \item Oxidized metal (MeO)
\end{enumerate}
Here, Met represents metal (consisting of iron Fe, iron-sulfide FeS and nickel
Ni) that will enter the core upon planetary melting due to their higher density
than the silicates. Sil represents the silicates (consisting of SiO$_2$ and MgO)
with lower density and higher melting points than metal. Wat represents water, a
major planetary building block beyond the ice line. Using the standard solar
abundance values from \cite{Lodders2003}, we use normalized metal and silicate
mass fractions $f_{\rm Met}=0.38$ and $f_{\rm Sil}=0.62$. We assume further that
all carbon in the protoplanetary disc binds with oxygen to form the
ultravolatile molecule CO and that all Si atoms bind with two O and all Mg atoms
with one O. This yields a surplus oxygen abundance to produce water with an
additional mass fraction $f_{\rm Wat}=0.38$, for a total mass of 1.38 of
MetSilWat when normalized to the sum of metal and silicates.

Clay (Cla) is simplified as an attachment of a water molecule to a silicate
mineral; we assume 15\% water mass fraction in clay minerals with an Mg:Si ratio
of approximately one \citep{LangeAhrens1982}. Considering clay as a single
species is a simplification \citep{Wilson+etal2006}, but this approach is
designed to capture in the simplest fashion the main storage of water in
silicates as well as its desiccation upon further heating. We treat oxidized
metal (MeO) by attaching oxygen atoms to the available iron and nickel atoms to
form FeO$_{4/3}$ and NiO after the ice melts (see discussion in Section
\ref{s:material_transitions}). Iron atoms bound with sulfur (FeS) are assumed to
be protected from oxidation and hence provide a baseline amount of core-forming
Met material even when the water abundance is high enough to oxidize all metal.
\begin{figure*}
  \begin{center}
    \includegraphics[width=0.9\linewidth]{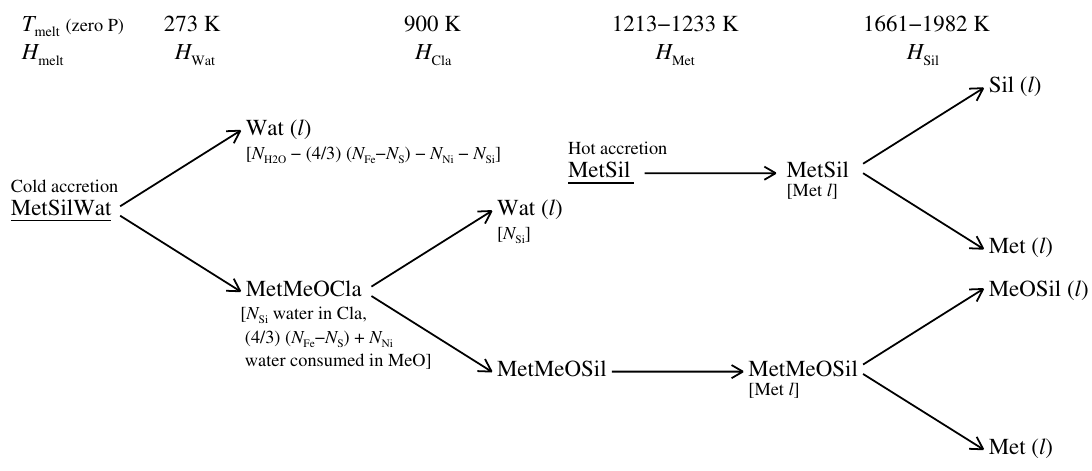}
  \end{center}
  \caption{Overview of the materials of the  ADAP code and their transitions.
  The accreted material is either MetSilWat (cold accretion) or MetSil (hot
  accretion where water vapour is lost in the accretion process). The code
  includes four material transitions: melting of water, desiccation of clay,
  melting of iron and melting of silicates. Liquid water is released both when
  ice is melted and when clays become unstable; the remaining water is
  transferred to oxidized metal MeO and clays.}
  \label{f:ADAP_structure}
\end{figure*}

\subsection{Material transitions}
\label{s:material_transitions}

We show an overview of how ADAP treats materials and their transitions in Figure
\ref{f:ADAP_structure}. The three fundamental species (Met, Sil and Wat) combine
into the primitive starting material MetSilWat. We assume that this material
melts in several distinct stages. At the melting temperature of water
($T=273.16$ K), the primitive and volatile-rich material MetSilWat splits into
Wat and MetMeOCla, with the relative fractions of siderophile Met and lithophile
MeO set by the available sulfur \citep[we set S/Fe $\approx$ 0.5 by number,
following the solar composition of][]{Lodders2003}. The surplus water separates
into a pure water annulus containing all the water molecules that were not
consumed by clay formation and metal oxidation.

Metallic iron and metallic nickel dissolve readily in water to form ferrous
hydroxide ${\rm Fe}{\rm (OH)}_2$ and nickel hydroxide ${\rm Ni} {\rm (OH)}_2$.
Here iron and nickel are doubly ionized (Fe$^{2+}$ and Ni$^{2+}$). Further
oxidation of ${\rm Fe}^{2+}$ to ${\rm Fe}^{3+}$ occurs at temperatures
above $100^\circ\,{\rm C}$ \citep{Fattah-alhosseini+etal2016}, precipitating out
as insoluble magnetite ${\rm Fe}_3{\rm O}_4$ by the Schikorr reaction
\begin{equation}
  3 {\rm Fe}{\rm (OH)}_2 \rightarrow  {\rm Fe}_3{\rm O}_4 + {\rm H}_2 + 2 {\rm
  H}_2{\rm O} \, .
  \label{eq:schikorr}
\end{equation}
We can consider CI chondrites as samples of a minor body that oxidized its iron
during aqueous alteration. Iron in the CI chondrites have an average valence
state of 2.77 \citep{Sutton+etal2017}. This corresponds approximately to
magnetite that has a mean valence state of $8/3 \approx 2.67$. We therefore bind
$4/3$ water molecules to each iron above the melting temperature of water to
mimic the formation of magnetite. The CI chondrites are part of the CC
(carbonaceous chondrite) group of meteorites, but we note that evidence of
oxidation of metallic iron by water has also been found in the ureilites from
the NC (non-carbonaceous) group, from the measured correlation between oxidized
iron and enrichment in the two heavy isotopes of oxygen that are associated with
water \citep{Sanders+etal2017}. Nickel has only two valence electrons and hence
oxidation can not proceed beyond Ni$^{2+}$.

The clay is assumed to desiccate at 900 K, releasing its (super-critical) water
contents. We use here a characteristic desiccation temperature for the clay
mineral serpentine; see \cite{LangeAhrens1982} for a discussion of the possible
range of desiccation temperatures for a range of clay minerals. The remaining
solid has the composition MetMeOSil.

The next melting event is the Fe-Ni-FeS mixture which is assumed to have a
solidus (0\% melting) and liquidus (100\% melting) temperature following
\begin{equation}
  T_{\rm melt} = T_0 \left( 1 + \frac{P}{P_0} \right)^q \, ,
  \label{eq:Tmelt}
\end{equation}
with $T_0=1213\,{\rm K}$ for the solidus and $T_0=1233\,{\rm K}$ for the
liquidus of Fe-Ni-FeS, $P_0=2 \times 10^{10}\,{\rm Pa}$ and $q=0.36$
\citep{SahijpalBhatia2015}. We assume here that the metallic melt does not
percolate to form a core until the silicates have melted significantly
\citep[see discussions on the efficiency of percolation in][]{Monteux+etal2018}.
However, we do include the latent heat of the melting metal (see Section
\ref{s:cp_L}).

The metal and the silicates finally split from the MetMeOSil at the silicate
melting temperature to form liquid Met and liquid MeOSil. We assume that MeO is
minerally bound to the silicates and therefore remains in the mantle. For the
melting temperature of Sil we consider two branches \citep{Monteux+etal2016}.
For low pressures, $P<20$ GPa, we use equation (\ref{eq:Tmelt}) with
$T_0=1661.2\,{\rm K}$, $P_0=1.336\,{\rm GPa}$ and $q=0.134$ for the solidus and
$T_0=1982.2\,{\rm K}$, $P_0=6.594\,{\rm GPa}$ and $q=0.186$ for the liquidus.
For $P>20$ GPa we use instead $T_0=2081.8\,{\rm K}$, $P_0=101.69\,{\rm GPa}$ and
$q=0.817$ for the solidus and $T_0=2006.8\,{\rm K}$, $P_0=34.65\,{\rm GPa}$ and
$q=0.542$ for the liquidus.

After the temperature in the envelope of the planet reaches the sublimation
temperature of water, which is the case for protoplanet masses above
approximately $0.01\,M_{\rm E}$ \citep{Johansen+etal2021}, we no longer allow
the planet to accrete water.  Instead the basic accreted material becomes
MetSil, which separates to Met and Sil at the silicate melting temperature. The
mantle of the final planet thus consists of the two components FeOSil and Sil.

\subsection{Material densities}

For the five basic material densities we choose:
\begin{enumerate}
  \item $\rho_{\rm Met}=5.7 \times 10^3\,{\rm kg\,m^{-3}}$ (Fe+Ni+FeS) or \\
  $\rho_{\rm Met}=4.8 \times 10^3\,{\rm kg\,m^{-3}}$ (pure FeS after melting
  MetSilWat)
  \item $\rho_{\rm Sil}=3.4 \times 10^3\,{\rm kg\,m^{-3}}$
  \item $\rho_{\rm Wat}=1.0 \times 10^3\,{\rm kg\,m^{-3}}$
  \item $\rho_{\rm Cla}=2.5 \times 10^3\,{\rm kg\,m^{-3}}$ (Sil+H$_2$O)
  \item $\rho_{\rm MeO}=5.8 \times 10^3\,{\rm kg\,m^{-3}}$ (FeO+NiO)
\end{enumerate}
Here the low density of Met reflects its high contents of sulfur
\citep{Morard+etal2018,Johnson+etal2020} and the density of Sil is based on the
uncompressed density of iron-depleted mantle material
\citep{Elkins-TantonSeager2008}. We assume for simplicity that the densities are
constant irrespective of temperature and pressure. The densities of composite
layers made out of these five base materials are constructed using volume
partitioning,
\begin{equation}
  \rho = \frac{1}{\sum f_i/\rho_i} \, ,
  \label{eq:rho}
\end{equation}
where $f_i$ is the normalized mass fraction, with $\sum_i f_i \equiv 1$ for each
layer. Clay contains 15\% by mass; we include here the full mass of the H$_2$
molecule and ignore any outgassing of hydrogen in the clay formation process.
The formation of oxidized metal MeO, on the other hand, is associated with
hydrogen loss. We do not keep track of this lost hydrogen since clay formation
happens in the earliest stages of planetary growth where the protoplanet cannot
yet hold onto an outgassed atmosphere.

\subsection{Differentiation}

We require that the material density decreases outwards from the centre for
stability. If a denser layer lies on top of a lighter layer, then we swap the
two layers. We calculate the new gravitational binding energy after the swap and
add the binding energy difference to the thermal energy of the two layers. We
require the two layers to maintain a shared temperature after the addition of
the thermal energy. This approach conserves total energy (gravitational plus
thermal) in the differentiation process.

\subsection{Heat transport}

We solve for the total energy $e$ present in each spherical shell. We choose to
have $e$ as the evolved variable because this eases the division of the
total energy between latent heat and temperature (see Section \ref{s:cp_L}). We
numerically solve the heat transfer equation 
\begin{equation}
  \mathcal{F} = -K \frac{\partial T}{\partial r} \, .
  \label{eq:Fheat}
\end{equation}
Here $\mathcal{F}$ is the heat flux and $K$ is the heat conductivity. We follow
\cite{Desch+etal2009} and discretize the heat transport across the interface
between shell $i$ and shell $i+1$ in an upwinded fashion,
\begin{equation}
  \mathcal{F}_i = -0.5 (K_i+K_{i+1}) \frac{T_{i+1}-T_i}{\delta r} \, .
\end{equation}
Here $\delta r$ is the radial grid size. We consider the mean of the heat
conductivities across the interface to avoid building up spuriously large
temperature gradients at interfaces between liquid and solid layers. The
evolution equation for the energy of cell $i$, $e_i$, becomes
\begin{equation}
  \dot{e}_i = 4 \pi r_{i-1}^2 \mathcal{F}_{i-1} - 4 \pi r_i^2 \mathcal{F}_i \, .
  \label{eq:edot}
\end{equation}

\subsection{Heat capacities and latent heat}
\label{s:cp_L}

We translate the energy in a cell $e_i$ to temperature $T_i$ taking into account
both the heat capacity and the latent heats. Heat capacities and latent heats
are assumed to be constant, independent of pressure and temperature. This way
the translation from energy to temperature can be calculated quickly and
precisely using analytical formulae. We use the NIST Chemistry Handbook
\citep{LinstromMallard2021} and \cite{LangeAhrens1982} to identify approximate
values for the heat capacity relevant at high temperatures:
\begin{enumerate}
  \item $c_{\rm Met} = 800\,{\rm J\,kg^{-1}\,K^{-1}}$
  \item $c_{\rm Sil} = 1200\,{\rm J\,kg^{-1}\,K^{-1}}$
  \item $c_{\rm Wat} = 4200\,{\rm J\,kg^{-1}\,K^{-1}}$
  \item $c_{\rm Cla} = 875\,{\rm J\,kg^{-1}\,K^{-1}}$
  \item $c_{\rm MeO} = 800\,{\rm J\,kg^{-1}\,K^{-1}}$
\end{enumerate}
We ignore the increase in $c_{\rm p}$ as the temperature reaches the critical
temperature for the material. The combined heat capacities of composite
materials are calculated from
\begin{equation}
  c_{\rm p} = \sum_i f_i c_i
\end{equation}
where $f_i$ are the mass fractions and $\sum f_i = 1$.

The latent heats of melting of the basic materials are assumed to take the
constant values \citep{Sahijpal+etal2007}
\begin{enumerate}
  \item $H_{\rm Met} = 270 \times 10^3 \,{\rm J\,kg^{-1}}$
  \item $H_{\rm Sil} = 400 \times 10^3 \,{\rm J\,kg^{-1}}$
  \item $H_{\rm Wat} = 334 \times 10^3 \,{\rm J\,kg^{-1}}$
\end{enumerate}
We ignore heat consumption in clay desiccation and set $H_{\rm Cla}=0$. The
latent heats of the composite materials are calculated from
\begin{equation}
  L = f_j H_j
\end{equation}
where $j=1,2,3$ represents Met, Sil and Wat and $f_j$ their total mass fractions
in a composite material. In this approach, the O in MeO is inherited from
oxidation of Fe and Ni by water and hence the translation of the energy of a
liquid MeO layer to temperature must, for consistency, take into account the
latent heat of both iron and water. Similar considerations apply to the Cla
material component.

\subsection{Thermal expansion coefficient}

The coefficient of thermal expansion is defined as
\begin{equation}
  \alpha = \frac{1}{V} \left( \frac{\partial V}{\partial T} \right)_P =
  \frac{1}{V} \sum_i \left( \frac{\partial V_i}{\partial T} \right)_P = \sum_i
  \frac{V_i}{V} \alpha_i \, ,
\end{equation}
where $V$ denotes the volume.  The coefficient enters calculations of the
adiabat and the convective heat conduction. For the basic materials we take
pressure and temperature-dependent parametrizations for $\alpha_{\rm Met}$ from
\cite{Chen+etal2007}, $\alpha_{\rm Sil}$ from \cite{Abe1997} and $\alpha_{\rm
Wat}$ for ice from \cite{Desch+etal2009} and for water from the IAPWS database
(see Paper II).

\subsection{Conduction and convection}

Equation (\ref{eq:Fheat}) describes heat transport by conduction with
conductivity $K$. We take approximate heat conductivity values from
\cite{Desch+etal2009} and \cite{Neumann+etal2012}:
\begin{enumerate}
  \item $K_{\rm Met} = 20\,{\rm W\,m^{-1}\,K^{-1}}$
  \item $K_{\rm Sil} = 4 \,{\rm W\,m^{-1}\,K^{-1}}$
  \item $K_{\rm Wat} = 1 \,{\rm W\,m^{-1}\,K^{-1}}$
\end{enumerate}
The heat conductivity of mixed layers is calculated using the geometric mean
model \citep{Neumann+etal2012}
\begin{equation}
  K = \prod_i K_i^{\theta_i} \, ,
\end{equation}
where $\theta_i = V_i/V = f_i (\rho/\rho_i)$ is the volume fraction of component
$i$, $f_i$ is the mass fraction and $\rho_i$ is the component density.

Convection is treated as an increased effective heat conduction
\citep{HeveySanders2006}. The Nusselt number, defined as
\begin{equation}
  {\rm Nu} = \frac{h L}{K}
\end{equation}
with $h = q/\Delta T$ denoting the energy flux $q$ divided by the temperature
difference $\Delta T$ and $L$ the length-scale of the layer, describes the
effective heat conduction due to convection relative to the microscopic heat
conduction, $K$. We consider convection to be a localized increased heat
transport with conductivity $K'$. We can thus write the Nusselt number as
\begin{equation}
  {\rm Nu} = \frac{K'}{K} \, .
\end{equation}
The effective heat conductivity depends on the melting degree $\varPhi$ defined
by
\begin{equation}
  \varPhi =
      \begin{cases}
      0 & \text{for $T \le T_{\rm sol}$} \\
      \frac{T-T_{\rm sol}}{T_{\rm liq}-T_{\rm sol}} & \text{for $T_{\rm sol} < T < T_{\rm liq}$} \\
      1 & \text{for $T \ge T_{\rm liq}$}
    \end{cases}
  \, .
\end{equation}
The solidus temperature $T_{\rm sol}$ and liquidus temperature $T_{\rm liq}$ are
composition dependent and follow equation (\ref{eq:Tmelt}).

Computer simulations of convection can be used to connect the Nusselt number to
the Rayleigh number ${\rm Ra}$, which describes the stability criterion for
convection. The Rayleigh number ${\rm Ra}$ is defined as
\begin{equation}
  {\rm Ra} = \frac{\alpha c_{\rm p} \rho^2 g (\Delta T) L^3}{K \eta} \, ,
\end{equation}
where $\alpha$ is the coefficient of thermal expansion, $c_{\rm p}$ is the heat
capacity of the fluid, $\rho$ is the density, $g$ is the gravitational
acceleration, $\Delta T$ is the drop in potential temperature over the
convective layer (i.e.\ the temperature drop relative to the adiabatic
temperature drop), $L$ is the length-scale of the layer, $K$ is the heat
conduction coefficient and $\eta$ is the dynamic viscosity of the fluid. We
follow \cite{Monteux+etal2016} and define the dynamic viscosity between the
solid end member value $\eta_{\rm s}$ and the liquid end member value $\eta_{\rm
m}$ as
\begin{equation}
  \eta = {\rm min} \left[ (1-\varPhi) \eta_{\rm s} + \varPhi \eta_{\rm m},
  \eta_{\rm m} \left\{ \frac{1}{(1-A) (1-\varPhi) + \varPhi} \right\}^{2.5}
  \right] \, .
\end{equation}
However, this expression is simplified compared to \cite{Monteux+etal2016} in
that we do not include the difference between the solid and molten density. The
viscosity displays here an abrupt fall to $\eta_{\rm m}$ around a melting
degree of $\varPhi \approx 0.4$. We assume that the dynamic viscosity of the
solid state is a constant $\eta_{\rm s}=10^{30}\,{\rm Pa\,s}$ for all materials.
For the dynamical viscosity of the molten phase we take
\begin{enumerate}
  \item $\eta_{\rm m,Met} = 1.5 \times 10^{-2}\,{\rm Pa\,s}$
  \citep{deWijs+etal1998}
  \item $\eta_{\rm m,Sil} = 100\, {\rm Pa\,s}$ \citep{Monteux+etal2016}
  \item $\eta_{\rm m,Wat} = A \exp \left( B/T + C T + D T^2 \right)$
  \citep[with coefficients from][]{Reid+etal1987}
\end{enumerate}

\cite{Solomatov2015} and \cite{Monteux+etal2016} describe two branches to
connect ${\rm Nu}$ and ${\rm Ra}$. The soft turbulent regime is valid for
${\rm Ra} < 10^{19}$. Here the Nusselt number is
\begin{equation}
  {\rm Nu} = \left( \frac{\rm Ra}{{\rm Ra}_{\rm c}} \right)^{1/3} \, .
\end{equation}
The critical Rayleigh number ${\rm Ra}_{\rm c}$ is approximately 1,000. The
models of the asteroid Vesta analysed in this paper has Rayleigh numbers up to
approximately $10^{21}$. At Rayleigh numbers above ${\rm Ra}>10^{19}$ the
convective heat transport is reduced in the hard turbulence regime, with Nusselt
number
\begin{equation}
  {\rm Nu} = \left( \frac{\rm Ra}{{\rm Ra}_{\rm c}} \right)^{2/7}
\end{equation}
and ${\rm Ra}_{\rm c} = 200$. This yields maximal Nusselt numbers of around
${\rm Nu} \approx 2 \times 10^5$ for the Vesta model.

\subsection{Radioactive heating}

We add to the energy equation (equation \ref{eq:edot}) also the heating by decay
of $^{26}$Al,
\begin{equation}
  \dot{e}_{{\rm 26},i} = M_{\rm Sil} E_{26} A_{26} \exp[-t/\tau_{26}] \, .
\end{equation}
Here $E_{26}=3.12 \,{\rm MeV}$ is the energy released by each decay minus the
energy carried by neutrinos \citep{Castillo-Rogez+etal2009}, $A_{26}=f_{\rm Al}
f_{26} / (m_{26} \tau_{26})$ is the decay rate per kg of Sil material at $t=0$,
$m_{26}$ is the mass of the $^{26}$Al atom and $\tau_{26}=1.05\times10^6\,{\rm
yr}$ is the decay constant. We have normalized $A_{26}$ to the Sil component
only.  Thus $f_{\rm Al}$ is the fraction of aluminium in the silicate part the
solar composition; we calculate $f_{\rm Al}=0.022$ by normalizing first relative
to hydrogen and helium and then relative to the total silicate contents
\citep{Lodders2003}. We take a base value of $f_{26} = 5 \times 10^{-5}$, but
note that the initial amount of $^{26}$Al may have been lower in the terrestrial
planet forming region than in the CAI forming regions closer to the star
\citep{Larsen+etal2011,Connelly+etal2012,Schiller+etal2015}. This gives an
initial heating rate of $5 \times 10^{-7}\,{\rm W\,kg^{-1}}$ for the silicates.
By construction, the Met and Wat components have no radiogenic heating.
Normalized to MetSil the heating rate is lower, $3 \times 10^{-7}\,{\rm
W\,kg^{-1}}$. This is about 1.5 times the value considered by
\cite{HeveySanders2006} for CI chondrites. However, the CI chondrites have a
density of only $2.2 \times 10^3\,{\rm kg\,m^{-3}}$ and contain a significant
amount of oxidized metal and clay.

\subsection{Accretion of mass and energy}

ADAP allows the planetesimal to grow in mass by accretion and includes the
accretional heating of the surface. The mass accreted in each time-step is
accumulated until it reaches a high enough value to create an additional grid
shell on top of the existing planetesimal. This new cell inherits the
temperature of the underlying grid cell. We add in each time-step accretion
heat to the outermost cell,
\begin{equation}
  \dot{e}_{\rm acc} = \frac{G M \dot{M}}{R} \, .
\end{equation}
Here $G$ is the gravitational constant, $M$ is the mass of the protoplanet,
$\dot{M}$ is the growth rate and $R$ is the radius of the protoplanet. We
nevertheless ignore the accretion phase here in Paper I and we checked that
indeed the accretion energy is negligible for the small planetesimals considered
here. The accretion heat does become important for massive, accreting
protoplanets when released under the thermal blanketing by a dense outgassed
atmosphere \citep{MatsuiAbe1986a,MatsuiAbe1986b}. We discuss this further in
Paper II.

\subsection{Removal of outgassed atmosphere}

Planetesimals and terrestrial planets forming by pebble accretion grow and
evolve within the protoplanetary gas disc and hence the outgassed volatiles can
escape by gas drag for low protoplanet masses. The gas flow around the
protoplanet is dominated by the sub-Keplerian shear flow $\Delta v$ for low-mass
objects and by the Keplerian shear flow beyond the so-called transition mass
\citep{LambrechtsJohansen2012}. For protoplanets below this transition mass from
headwind flow to shear flow, the mass removal by the sub-Keplerian headwind
$\Delta v$ can be quantified as
\begin{equation}
  \dot{M}_{\rm atm} = \int_R^\infty \rho_{\rm atm} \Delta v (2 \pi r) {\rm d}r
  \, .
\end{equation}
Here $\rho_{\rm atm}$ is the density of the atmosphere of outgassed volatiles.
We assume now that $\rho_{\rm atm} = M_{\rm atm}/(4 \pi R^2 H_{\rm atm})$, where
$H_{\rm atm}$ is the scale-height of the outgassed atmosphere. That gives then
\begin{equation}
  \dot{M}_{\rm atm} \approx \frac{M_{\rm atm}}{4 \pi R^2 H_{\rm atm}} \Delta v
  (2 \pi R) H_{\rm atm} \, .
\end{equation}
We rewrite the mass loss in terms of the mass removal timescale,
\begin{equation}
  \dot{M}_{\rm atm} = \frac{M_{\rm atm} \Delta v}{2 R} = \frac{M_{\rm
  atm}}{\tau_{\rm rmv}} \, .
\end{equation}
Here the removal timescale is
\begin{equation}
  \tau_{\rm rmv} = \frac{2 R}{\Delta v} = 2 \times 10^4\,{\rm s}\,\left(
  \frac{R}{250\,{\rm km}} \right) \left( \frac{\Delta v}{25\,{\rm
  m\,s^{-1}}} \right)^{-1} \, .
  \label{eq:taurmv}
\end{equation}
This should now be compared to the timescale for outgassing via thermal
desorption,
\begin{equation}
  \dot{M}_{\rm atm}^{\rm (out)} = 4 \pi R^2 v_{\rm th} \rho_{\rm sat} \, .
\end{equation}
Here $\rho_{\rm sat}$ is the saturated vapour density which describes both the
equilibrium vapour density above a solid surface as well as the outgassing rate
in the absence of a bound atmosphere. We consider here only outgassing by
sublimation or evaporation of volatiles from the surface of planetesimals and
small protoplanets; larger bodies experience instead outgassing of volatiles
from the magma ocean but these protoplanets will be massive enough to retain
their outgassed atmosphere. We rewrite the thermal outgassing timescale using
the ideal gas law as
\begin{equation}
  \dot{M}_{\rm atm}^{\rm (out)} = 4 \pi R^2 v_{\rm th} \frac{\mu_{\rm atm}
  P_{\rm sat}}{k_{\rm B} T} = \frac{\mu_{\rm atm} g M_{\rm sat} v_{\rm
  th}}{k_{\rm B} T} \, .
\end{equation}
Here we made use of the fact that the surface pressure and the atmospheric
mass are connected through $P_{\rm sat} = g M_{\rm sat}/(4 \pi R^2)$ with $g = G
M/R^2 = (4 \pi/3) G \rho_\bullet R$ denoting the gravitational acceleration at
the surface ($\rho_\bullet$ is the internal density of the protoplanet). The
mass of the saturated atmosphere, $M_{\rm sat}$, is the maximal atmosphere mass
that can be sustained by thermal outgassing. The outgassing timescale is thus
\begin{eqnarray}
  \tau_{\rm out} &=& \frac{M_{\rm sat}}{\dot{M}_{\rm atm}^{\rm (out)}} = \frac{k_{\rm B} T}{\mu_{\rm atm} g v_{\rm th}} =
  \frac{v_{\rm th}}{g} \nonumber \\
  &=& 2 \times 10^3\,{\rm s} \left( \frac{T}{160\,{\rm K}}
  \right)^{1/2} \left( \frac{\rho_\bullet}{2 \times 10^3\,{\rm kg\,m^{-3}}}
  \right)^{-1} \times \nonumber \\
  && \left( \frac{\mu_{\rm atm}}{\mu_{\rm H_2O}} \right)^{-1/2} \left(
  \frac{R}{250\,{\rm km}} \right)^{-1} \, .
\end{eqnarray}
Here $v_{\rm th} = \sqrt{k_{\rm B} T/\mu_{\rm atm}}$ is the thermal speed of the
molecules, here scaled to the mass of the water molecule. Comparing this
expression to the mass loss timescale of equation (\ref{eq:taurmv}), we see
that the hydrodynamical timescale is longer than the outgassing timescale for
all relevant $T$ and $R$. We therefore remove the outgassed atmosphere on the
hydrodynamical timescale when the protoplanet is below the transition mass
between headwind flow and Keplerian shear flow. The transition mass is derived
by \cite{LambrechtsJohansen2012} to be \begin{equation}
  M_{\rm t} = \sqrt{\frac{1}{3}} \frac{\Delta v^3}{G \varOmega} \, .
\end{equation}
Here $\varOmega = \sqrt{G M_\star/r^3}$ is the Keplerian frequency at distance
$r$ from the central star of mass $M_\star$. The transition mass is
approximately $M_{\rm t}=10^{-4}\,M_{\rm E}$ for a nominal protoplanetary disc
at 1 AU with $\Delta v = 25\,{\rm m\,s^{-1}}$. We hence remove the outgassed
atmosphere in our model when the planetary mass is below $10^{-4}\,M_{\rm E}$.
\begin{figure}
  \begin{center}
    \includegraphics[width=0.9\linewidth]{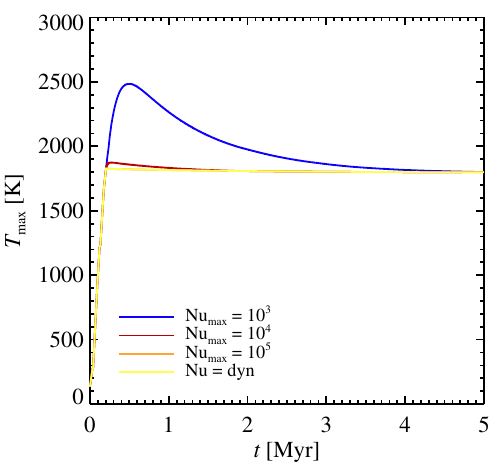}
  \end{center}
  \caption{Plot of the maximum temperature within a planetesimal of radius 250
  km as a function of time for three fixed values of the Nusselt number and a
  dynamical setting of Nu. Fixing Nu to a too low value clearly leads to an
  overestimate of the interior temperature. We therefore run the simulations
  with a dynamical value of Nu that is calculated based on the Rayleigh number
  of each convective layer.}
  \label{f:Tmax_t}
\end{figure}
\begin{figure*}
  \begin{center}
    \includegraphics[width=0.9\linewidth]{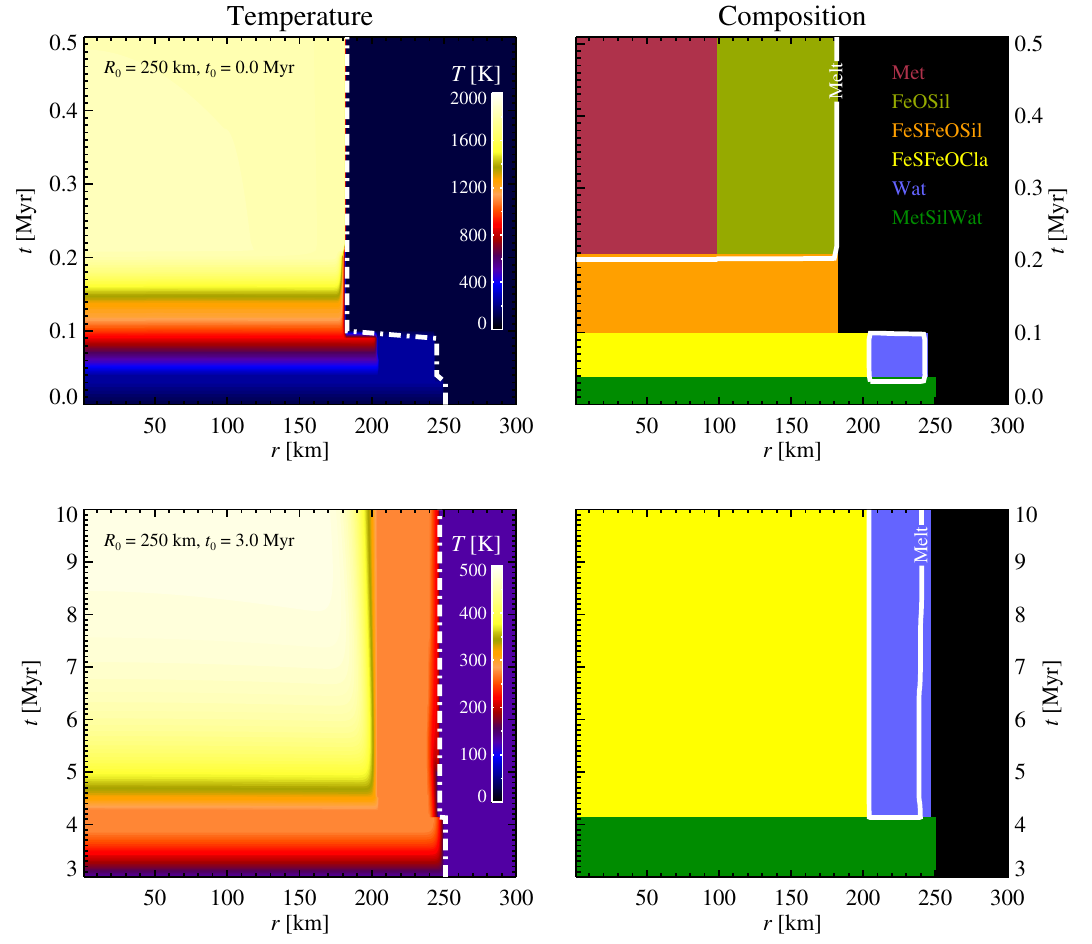}
  \end{center}
  \caption{Evolution of the interior temperature (left) and the composition of
  the layers (right) as a function of distance from the planetesimal centre and
  time. The top panels show the results when starting at $t_0 = 0$ (assuming the
  maximal possible inventory of $^{26}$Al) and the bottom panels when starting
  at $t_0 = 3$
  Myr (four $^{26}$Al half-lives later). The early-formed Vesta analogue heats
  rapidly to an interior temperature of 2,000 K within 200,000 years. The
  composition changes from primitive MetSilWat to FeSFeOCla at the melting
  temperature of water. The clay desiccates to form FeSFeOSil at 900 K. Finally,
  the FeS separates from the mantle material at the melting temperature of the
  silicates, to form an internally differentiated planetesimal with a high FeO
  content in the mantle as the only memory of its icy origins. The Ceres
  analogue starting after 3 Myr never heats above the desiccation temperature of
  clay. The clay interior is surrounded by a slowly freezing water ocean
  containing the excess water that could not enter the silicate minerals.}
  \label{f:planetesimal_structure_250km}
\end{figure*}

\section{Heating and differentiation of planetesimals}

We consider Vesta as a prototype of the population of protoplanets
\citep{Russell+etal2012} that formed in the terrestrial planet region and
competed for pebbles to grow to planetary sizes. Vesta is differentiated and has
an iron core of approximately 110 km in radius, giving a core mass fraction of
approximately 18\% \citep{Russell+etal2012}. Its thermal evolution may be
connected to that of Ceres, which remained volatile-rich and avoided core
formation, with the main difference being the formation time and hence the
amount of radiogenic heating available \citep{McCordSotin2005}.  The mantle of
Vesta contains a large fraction of oxidized iron FeO, approximately 24\%
inferred from HED meteorites, observations with the Dawn spacecraft and
comparisons with the H-chondrites \citep{Toplis+etal2013,Tronnes+etal2019}. We
take the high oxidation as evidence that Vesta formed exterior of the ice line
with a substantial water fraction. This ice then formed clays and oxidized
metallic iron when it melted, leaving the core to form mainly from FeS, which is
siderophile and provides the main sulfur carrier at high temperatures
\citep{Scott+etal2002}. The clays would have subsequently dried out after
heating above 900 K, leaving water only as structurally bound remnants in
apatite minerals \citep{Sarafian+etal2014}.

\subsection{Planetesimal model setup}

To facilitate comparisons with Vesta data and to differences between Vesta and
Ceres, we fix the planetesimal radius to $R=250\,{\rm km}$ and the background
temperature of the protoplanetary disc is set to $T_0=135\,{\rm K}$.  The radial
resolution element is chosen to be $\delta r = 1\,{\rm km}$. The initial
composition is given in the relative proportions $({\rm Met},{\rm Sil},{\rm
Wat}) = (0.38, 0.62, 0.38)$. This yields a total water mass fraction of 28\%
relative to the sum of metal and silicates. The metal material includes here
sulfur in FeS and its fraction is therefore larger than the core mass fraction
of Earth, which likely contains far less than the solar abundance of sulfur. The
sum of Met, Sil and Wat forms the primitive MetSilWat material with a density of
approximately 2,000 ${\rm kg\,m^{-3}}$. We ignore for simplicity the porosity
evolution of the planetesimal as well as the possibility that Al is transferred
to the first melts and hence concentrated near the surface
\citep{Neumann+etal2012,Golabek+etal2014}.

We start by testing the robustness of the numerical scheme and our treatment of
convection in 1-D.  In Figure \ref{f:Tmax_t} we show the maximum temperature
within the planetesimal as a function of time for a simulation with a dynamical
calculation of the Nusselt number Nu and three simulations where Nu is capped at
values of $10^3$, $10^4$ and $10^5$, respectively. Evolving the planetesimal at
an artificially capped Nu allows for a much longer time-step but at the expense
of an exaggerated temperature in the interior. We therefore opt to run the
simulations with a dynamical Nusselt number. For the planetesimal case, the
dynamical Nusselt number of the water and magma layers is approximately $10^5$,
but for larger planets considered in Paper II and Paper III the Nusselt number
reaches values 2-3 orders of magnitude larger.

\subsection{Interior temperature and composition of planetesimals}

We show space-time plots of the results of our heating and differentiation model
in Figure \ref{f:planetesimal_structure_250km} for two different starting times.
Our planetesimals start with a significant fraction of water mixed with metal
and silicates (MetSilWat). As the water melts at the triple point of H$_2$O, the
ice component transforms to phyllosilicates, oxidized iron and a surface water
layer. The water layer is expelled immediately due to its high temperature in
case of the early-formed Vesta analogue. Troilite FeS remains with the silicates
and is the main core forming material since it is protected from oxidation. As
the temperature reaches 900 K, the water in the clay is released and expelled
from the planetesimal. The iron sulfide melts at 1213 K but does not separate
from the silicates before the silicates melt at approximately 1700 K. This leads
to the formation of an FeS core surrounded by a mantle of silicates and oxidized
iron. The results of the early-formed planetesimal agree well with other Vesta
models \citep{GhoshMcSween1998,GrimmMcSween1989,GrimmMcSween1993}, although we
demonstrate here how the high FeO contents of Vesta are likely not a
consequence of any increased oxidation of iron with distance from the Sun
\citep{RighterO'Brien2011}, but rather a fingerprint of the accretion of Vesta
outside of the water ice line with a substantial amount of ice.
\begin{figure}
  \includegraphics[width=0.9\linewidth]{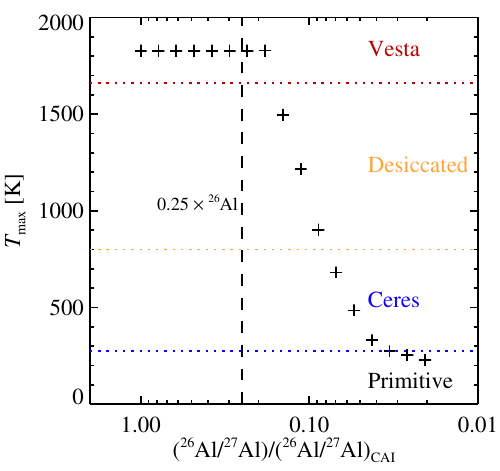}
  \caption{Maximum temperature of planetesimals of 250 km in radius for
  different values of the $^{26}$Al/$^{27}$Al ratio relative to that of the
  CAI-forming region. For early starting times, within approximately two
  $^{26}$Al half-lives ($t_{1/2} = 700,000\,{\rm yr}$), the resulting structure
  becomes like Vesta with an iron core and a mantle with a high content of FeO.
  Starting times between two and three half-lives instead lead to the formation
  of a desiccated body that has lost its minerally bound water but not
  differentiated. Ceres analogues consisting of clay and ice form between three
  and four half-lives and the original primitive composition can be maintained
  by an even later formation. We mark with a dashed line the likely
  reduced $^{26}$Al/$^{27}$Al budget in the asteroid belt relative to the
  CAI-forming region \citep{Larsen+etal2011,Schiller+etal2015}.}
  \label{f:Tmax_t0}
\end{figure}

The lower panels of Figure \ref{f:planetesimal_structure_250km} show the results
of forming the planetesimal later, at $t=3$ Myr or after approximately four
$^{26}$Al half-lives. Here the maximum temperature reached is 400 K and the
interior structure is dominated by a phyllosilicate 'core' surrounded by a
mantle of excess water that could not enter the clays. The water ocean
crystallizes slowly after the $^{26}$Al heat runs out.

\subsection{Varying the $^{26}$Al abundance}

In Figure \ref{f:Tmax_t0} we show the maximum temperature of the planetesimal
as a function of the starting ratio of $^{26}$Al/$^{27}$Al. The starting amount
of $^{26}$Al plays a decisive role in the final composition of the planetesimal.
Planetesimals that form within two half-lives of $^{26}$Al undergo the sequence
of melting and differentiation that happened to Vesta; the water is then only
present in the form of FeO in the final body. Forming in the interval between
two and three half-lives instead leads to the formation of a desiccated body
that lost its water but did not heat enough to differentiate into a silicate
mantle and a metal core. Formation in the interval between three and four
lifetimes gives maximum temperatures below the desiccation temperature of clay.
These Ceres analogues are dominated in their interiors by clay and the mantle
consists of the excess water that could not enter the phyllosilicates. Finally,
primitive bodies akin to comets and Kuiper belt objects form after four
half-lives of $^{26}$Al.

\subsection{Mapping the thermal evolution of planetesimals}

\begin{figure}
  \includegraphics[width=0.9\linewidth]{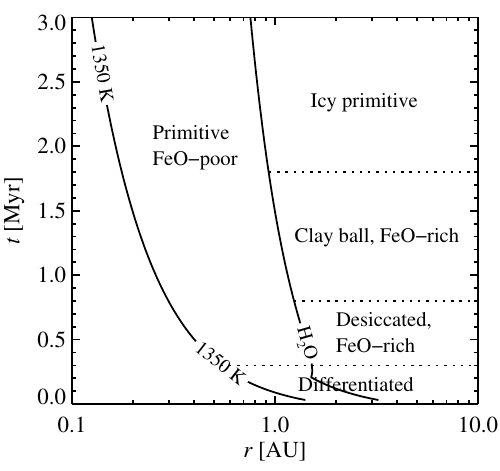}
  \caption{Map of the formation region of planetesimals as a function of time.
  The position of the water ice line and the silicate sublimation line (1350 K)
  are calculated from the dead zone model of \cite{Johansen+etal2021} that
  considers stellar irradiation as well as viscous heating where the thermal
  ionization level is above the threshold for the magnetorotational instability.
  We assumed here a basic $^{26}$Al/$^{27}$Al of 0.25 times the value in the
  CAI-forming region \citep{Larsen+etal2011,Connelly+etal2012}. Exterior of the
  water ice line, FeO-rich bodies form when the formation time is early enough
  to melt the ice. Differentiated and primitive FeO-poor bodies form interior of
  the water ice line, perhaps analogues of the enstatite chondrites.}
  \label{f:planetesimal_map}
\end{figure}

The maximum interior temperature and the initial composition can be mapped onto
the thermal evolution of the inner regions of the solar protoplanetary disc. We
adopt the protoplanetary model of \cite{Johansen+etal2021}, which considers
the irradiation from the forming star as well as viscous heating close to the
star where the ionization degree is high enough to trigger the magnetorotational
instability \citep{DeschTurner2015}. We take into account here that the initial
$^{26}$Al/$^{27}$Al ratio in the formation region of asteroids in the Solar
System appears to have been lower outside of the CAI-forming region
\citep{Larsen+etal2011,Connelly+etal2012}; we thus use 1/4 of the CAI value to
be consistent with these measurements \citep{Schiller+etal2015}. This yields the
planetesimal map displayed in Figure \ref{f:planetesimal_map}. Exterior of the
water ice line, FeO-rich planetesimals form by melting of the ice, unless the
planetesimal forms after 1.8 Myr or so. Between the silicate sublimation front
and the water ice line, we have FeO-poor bodies akin to enstatite chondrites or
aubrites \citep{Keil2010}.

\section{Core mass fraction and FeO fraction in the mantle}
\label{s:fcore_fFeO}

We will present full models of heating and differentiation of terrestrial
planets forming by pebble accretion in Paper II and Paper III. Here we show an
important intermediate result, namely that the core mass fraction and FeO mantle
fraction of planets can be generalized directly from the planetesimal heating
model presented in the previous section.

\subsection{Core and mantle iron in Solar System bodies}

The distribution of iron between metal core and mantle FeO within a
differentiated body carries a history of the amount of water that was accreted.
The fraction of FeO in the mantle is known for Venus, Earth, Mars and Vesta. The
FeO fraction increases with distance from the Sun \citep{RighterO'Brien2011} and
this increase is traditionally ascribed to an increasing oxidation state of the
dust with increasing distance from the star. The high FeO fraction in the
matrices of carbonaceous chondrites is taken as support for this picture
\citep{Rubie+etal2015}. However, the carbonaceous chondrites likely accreted
with significant ice fractions and experienced extensive aqueous alteration;
hence the primordial oxidation state of the iron in the pebbles that orbited
around the young Sun is not known.

\begin{figure}
  \includegraphics[width=0.9\linewidth]{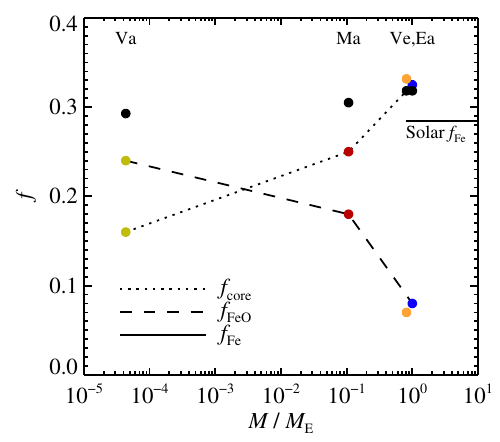}
  \caption{Measured values for core mass fraction (dotted line), FeO mass
  fraction in the mantle (dashed line) and the total iron contents (black
  circles) of Vesta (Va), Mars (Ma), Earth (Ea), Venus (Ve) and the Sun. All
  these measurements come with uncertainties that are not plotted here;
  references for the data are given in the main text. The measured core mass
  fraction clearly increases with increasing mass, while the measured FeO mantle
  fraction decreases accordingly.}
  \label{f:fFeO_fcore}
\end{figure}

We plot the core mass fraction and the FeO mantle fraction of Vesta, Mars, Earth
and Venus in Figure \ref{f:fFeO_fcore}. The core size of Vesta is constrained to
be approximately 110 km from the gravitational moment measured by Dawn and by
geochemical measurements of HED meteorites
\citep{Ruzicka+etal1997,RighterDrake1997,Russell+etal2012}. This gives a core
mass fraction of approximately 18\% of the total mass of Vesta. The FeO mantle
fraction is inferred to be 24\% from HED meteorites that originate from Vesta
\citep{Toplis+etal2013,Tronnes+etal2019}. For Mars, we take a core mass
fraction of 25\% \citep{Rivoldini+etal2011,Stahler+etal2021} and the FeO mass
fraction in the mantle to be 18\% \citep{RobinsonTaylor2001}. For Earth, the
core mass is $1.9 \times 10^{24}\,{\rm kg}$, the core mass fraction is 32.5\%
and the FeO mass fraction in the mantle is 8\% \citep{RobinsonTaylor2001}. For
Venus, the core mass is not known but the FeO mass fraction in the mantle is
estimated between 6.5\% and 8.1\% from the Venera landers
\citep{Surkov+etal1984,Surkov+etal1986,RobinsonTaylor2001}. The total iron is
typically larger than the solar abundance of 28.4\% \citep{Lodders2003}; this
could be due to uncertainties in the measurements of core mass and FeO fraction
or loss of silicon and magnesium relative to iron prior to accretion
\citep{Hin+etal2017}. Figure \ref{f:fFeO_fcore} shows that the FeO mantle
fraction is a monotonically decreasing function of planetary mass, while the
core mass fraction is monotonically increasing.
% ZFe = 8.380d5*56/(2.431d10+2.343d9*4)
% Zsil = 0.00303
% ZNi = 4.780d4*59/(2.431d10+2.343d9*4)
% ZFe/(ZFe+Zsil+Zni) = 0.309 = 30.9%

\subsection{Light elements in the core}

When discussing core mass fractions, the possible presence of lighter elements
in the core becomes important to take into consideration. The density of the
Earth's core is consistent with incorporation of both Si and O into the metallic
melt during differentiation \citep{Badro+etal2015}. The cores of Vesta and Mars
could both contain approximately 15\% weight sulfur \citep{Rivoldini+etal2011}.
The sulfur weight fraction can be maximally 36\% if the core consists entirely
of FeS. Using the solar composition \citep{Lodders2003} we have ${\rm S}/{\rm
Fe} = 0.53$ in number, which would yield a core mass fraction of at least 27\%
if all S is transported to the core in the form of siderophile FeS. This is
clearly above the estimate for Vesta and Mars. Sulfur could nevertheless have
entered the core and expelled later due to its incompatibility with solid iron
\citep{Johnson+etal2020}. \cite{Steenstra+etal2019} infer 15\% sulfur in the
core of Vesta from chalcophile elements. This is consistent with measurements of
the original S contents of iron meteorites \citep{Chabot2004}, later expelled
due to incompatibility with the solid iron phase. We therefore consider 53\% of
the iron to have been originally in the form of FeS. This attachment is
important, since the iron is protected from oxidation by the sulfur. We
additionally consider 5\% Ni in the cores and furthermore 10\% light elements by
mass for Venus, Earth and Vesta. For Mars, based on seismic measurements by the
Insight mission, we take 5\% Ni and 15\% light elements in the core
\citep{Stahler+etal2021}.

\subsection{The sulfur content of chondrites}

The CI chondrites contain an approximately solar composition of S, with S/Fe =
0.5 -- all other meteorite classes are depleted relative to this value
\citep{Ebel2011}. FeS sublimates at approximately 700 K under nominal
protoplanetary disc abundances and pressures \citep{Lodders2003}.  This
temperature makes S volatile in the envelopes of the more massive planets. S
should therefore strongly record the temperature of the envelope of the parent
body, with asteroids accreting nominal values and planets successively missing
more S with increasing mass. Earth's mantle contains only approximately 200 ppm
of sulfur \citep{Jackson+etal2021}, while the core of Earth may hold significant
S due to metal-silicate partition coefficients in the range 100--1,000 for the
relevant pressures at the core-mantle boundary
\citep{Boujibar+etal2014,Jackson+etal2021}. Chondrules were already depleted in
sulfur \citep{MarrocchiLibourel2013} and hence accretion of thermally processed
chondrules constitutes an additional pathway to S depletion before accretion.

\subsection{Planet formation model}

We use the planet formation model of \cite{Johansen+etal2021} to test if the
pattern evident in Figure \ref{f:fFeO_fcore} is consistent with oxidation of
iron by interaction with water outside of the water ice line. This model
considers the formation and migration of terrestrial planets that accrete both
pebbles and planetesimals. The planetesimals are assumed to form within a ring
of width 0.1 AU at a distance of 2.3 AU from the host star, hence the
protoplanet can only accrete planetesimals until it migrates out of the ring.
The pebbles are assumed to drift through the protoplanetary disc. The
planetesimal contribution is high ($\sim$50\%) as a protoplanet grows up to the
mass of Mars; for higher masses pebble accretion strongly dominates.

We calculate for each time-step in the planet formation model of
\cite{Johansen+etal2021} the accreted iron mass, with the iron fraction based on
the total amount calculated per body in Figure \ref{f:fFeO_fcore} to take into
account a varying iron content in the accreted material for the different
bodies. We assume that the iron in the accreted pebbles has a low oxide fraction
(FeO, 5\%) and high sulfide fraction (FeS, 45\%), broadly consistent with
spectral analysis of interstellar dust \citep{Westphal+etal2019}\footnote{These
authors find oxide fraction of total iron $\zeta_{\rm FeO}<0.35$ and sulfide
fraction $\zeta_{\rm FeS}<0.45$, with the remaining fraction $\zeta_{\rm Fe(m)}$
in metallic iron form. These values are also consistent with the low FeO
contents of enstatite chondrites and H ordinary chondrites where the iron has
likely been reduced by chondrule formation in a reducing gas
\citep{Connolly+etal1994}. Even H chondrites, with their contents of metallic
iron, are overall iron poor compared to the solar composition and may have
preferentially accreted rocky chondrules over the metal nuggets that carry most
of the metallic iron.}. We assume that the iron accreted together with water
becomes 100\% oxidized. We furthermore assume that iron in the accreted
planetesimals is 100\% oxidized, except for the iron bound in FeS.  All the
protoplanets are started at $r_0=1.6$ AU.
\begin{figure*}
  \begin{center}
    \includegraphics[width=0.9\linewidth]{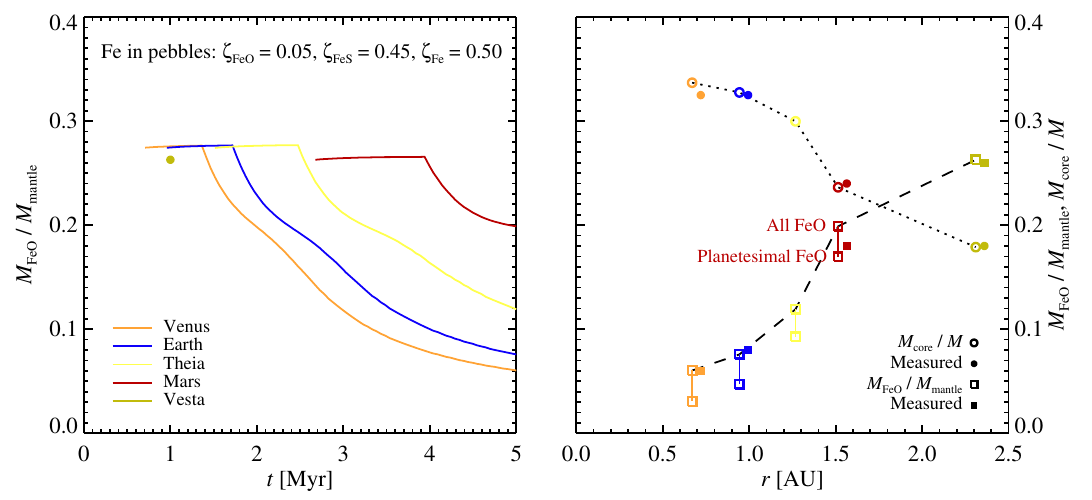}
  \end{center}
  \caption{The FeO fraction in the mantle as a function of time (left) and the
  final FeO fraction and core fraction (right). We assume here that 5\% of
  the iron in the pebbles is in FeO form and 45\% in FeS. The FeO fraction
  starts high as all iron that is not in FeS becomes oxidized by dissolution in
  water. The value drops after ice accretion terminates at around $0.01\,M_{\rm
  E}$.  The resulting core mass fraction and FeO mantle fraction yield good
  matches to measurements of Venus, Earth, Mars and Vesta, demonstrating that
  these bodies could share a common origin exterior of the water ice line where
  they initially accreted significant amounts of ice. The connected values of
  the FeO fraction of the model indicate the total FeO contribution from pebbles
  and planetesimals and the contribution from FeO-rich planetesimals alone.}
  \label{f:FeO_t}
\end{figure*}

\subsection{Oxidation state of Vesta}

The oxidation of iron by dissolution in hot water through the Schikorr reaction
(equation \ref{eq:schikorr}) should yield an average iron oxidation similar to
magnetite (Fe$_3$O$_4$), containing two Fe$^{3+}$ and one Fe$^{2+}$ to balance
the four O$^{2-}$. This would imply a strongly oxidizing mantle near the QFM
(Quartz-Fayalite-Magnetite) buffer \citep{Ortenzi+etal2020}. From the abundances
of siderophile elements in meteorites from Vesta, on the other hand, the
oxidation state seems to have been much lower, perhaps a logarithmic unit below
the IW (Iron-W\"ustite) buffer with very low Fe$^{3+}$ fractions
\citep{Righter+etal2016,Steenstra+etal2019}. This would seem to contrast with
our oxidation mechanism presented in Section 3. However, Vesta is a small body
and may have been effectively an open system in contact with the protoplanetary
gas disc. The removal timescale by gas flows for an outgassed atmosphere was
identified in equation (\ref{eq:taurmv}) to be a few hours only. The QFM
buffer has an oxygen fugacity (partial pressure of oxygen measured in bars) of
$f{\rm O}_2 \sim 10^{-8}$. The mass of an outgassed oxygen atmosphere is
\begin{equation}
  M_{\rm atm} = 4 \pi R^2 P / g \, ,
\end{equation}
where $R$ is the radius and $g$ is the surface gravity. This yields $M_{\rm
atm,O} = 3.6 \times 10^9\,{\rm kg}$ for Vesta at the QFM buffer. With a removal
timescale of $\tau_{\rm rmv} = 2 \times 10^4\,{\rm s}$ from equation
(\ref{eq:taurmv}), the oxygen removal rate is $\dot{M} \sim 5.6 \times
10^{18}\,{\rm kg\,Myr^{-1}}$.  The mass of Vesta is $M_{\rm Va} = 2.6 \times
10^{20}\,{\rm kg}$. With an iron mass fraction of $f_{\rm Fe}=0.284$, 55\% of
this iron oxidized to magnetite with 4/3 O for each Fe, we get a mass of $M \sim
1.7 \times 10^{19}\,{\rm kg}$ of oxygen bound with iron. The removal time of the
oxygen is thus only approximately 3 Myr. Once the mantle has been reduced enough
that Fe$^{3+}$ becomes scarce, then the oxygen fugacity is set by the
Iron-W\"ustite buffer instead, approximately four orders of magnitude lower than
the QFM buffer. Hence the removal timescale goes from millions of years to
billions of years and the oxygen is largely protected and the oxidation state is
maintained.  Motivated by these considerations, we therefore assume that the
magnetite formed by water flows on planetesimals is reduced to fayalite
(Fe$_2$SiO$_4$) through oxygen outgassing and loss to the protoplanetary
disc\footnote{We note that we do not take into account that the growing
protoplanets may accrete planetesimals that have not yet reduced their magnetite
to fayalite, nor do we include the possibility of increasing the FeO mass
fraction by transfer of oxygen from SiO$_2$ to metallic Fe
\citep{Rubie+etal2015} or the oxidation of FeO to FeO$_{3/2}$ accompanied by
reduction of another FeO to Fe at high pressures in the magma ocean
\citep{Armstrong+etal2019}.}.
\begin{figure}
  \begin{center}
    \includegraphics[width=0.9\linewidth]{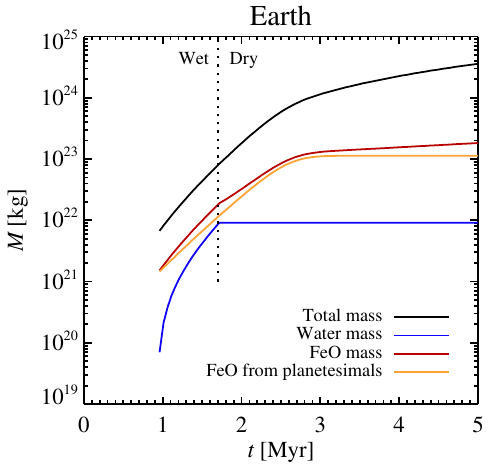}
  \end{center}
  \caption{The growth of Earth and the accretion of FeO in the pebble accretion
  model of \cite{Johansen+etal2021}. Water is accreted up until the protoplanet
  reaches $0.01\,M_{\rm E}$ after 1.6 Myr. The FeO mass increases initially
  because of oxidation of iron in the water as well as due to accretion of
  FeO-rich planetesimals (which were also oxidized by water flow). Planetesimal
  accretion then takes over as the main delivery of FeO, before the planet
  migrates out of the birth planetesimal belt. This results in a late FeO
  contribution that is dominated by pebbles drifting in from the outer Solar
  System composition, in agreement with the match of the iron isotopic
  composition of CI chondrites to iron in the Earth's mantle
  \citep{Schiller+etal2020}.}
  \label{f:mass_growth_Earth}
\end{figure}
\begin{figure*}
  \begin{center}
    \includegraphics[width=0.9\linewidth]{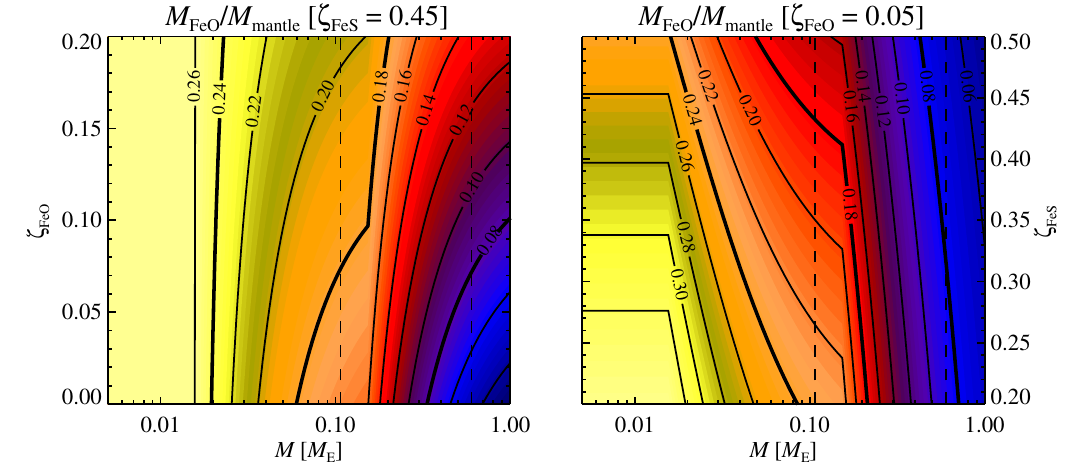}
  \end{center}
  \caption{The results of a simple model for calculating the final FeO fraction
  of planetary bodies as a function of the final mass. We maintain a constant
  $\zeta_{\rm FeS}=0.45$ in the left panel and vary the FeO fraction $\zeta_{\rm
  FeO}$ of the accreted pebbles, while in the right panel we fix $\zeta_{\rm
  FeO}=0.05$ and vary $\zeta_{\rm FeS}$.  The FeO mantle fractions of Vesta
  (24\%), Mars (18\%) and Earth (8\%) are indicated with thick contours and the
  planetary masses of Mars and proto-Earth (before the moon-forming planetary
  collision) are indicated with dashed lines. The FeO fraction of Earth is very
  sensitive to the FeO fraction of the accreted pebbles, with best fits obtained
  around $\zeta_{\rm FeO}=0.05$ . Mars, on the other hand, obtains most of its
  FeO from oxidizing its metallic Fe and accreting FeO-rich planetesimals and
  its mantle FeO contents are relatively independent of the oxidation degree of
  the pebbles. The FeO fraction of Vesta's mantle depends mainly on the FeS
  fraction of the pebbles. The right plot shows that low values of $\zeta_{\rm
  FeS}$ lead to very high mantle FeO fractions for low-mass planets and
  planetesimals.  The FeO mantle fractions of Mars and Earth are relatively
  independent of the FeS fraction, although values below $\zeta_{\rm FeS}=0.3$
  yield too low core masses and too high FeO mantle fractions.}
  \label{f:simple_FeO_model}
\end{figure*}

\subsection{Calculated FeO fractions}

In Figure \ref{f:FeO_t} we show the calculated FeO mantle fraction and core mass
fraction from the planet formation model of \cite{Johansen+etal2021}. We show in
the figure both the total FeO mantle fraction in the model and that of
planetesimals separately, to illustrate that particularly Mars has a large
contribution of FeO from accreting Vesta-like planetesimals. Overall, and
despite uncertainties in both model and data, we find an intriguing match
between model and measurements.  The model nevertheless overpredicts the FeO
fraction of the martian mantle by a few percent. This could indicate
either that our source material (pebbles and differentiated planetesimals) is
too rich in FeO or that some FeO has been transferred from the mantle to the
core after core formation \citep{Frost+etal2010,Davies+etal2020}.

We analyse the growth and accretion of FeO for our Earth analogue more in Figure
\ref{f:mass_growth_Earth} to support the interpretation of the results presented
in Figure \ref{f:FeO_t}. We ignore here, as in Figure \ref{f:FeO_t}, any
addition of FeO to the terrestrial mantle from the impactor in the moon-forming
impact. Earth starts its growth at a mass of $10^{-3}\,M_{\rm E}$ after 1 Myr.
The planet accretes initially together with significant amounts of ice, but the
gas envelope becomes hot enough to sublimate water after reaching $160$ K in the
envelope at a mass of approximately $0.01\,M_{\rm E}$.  The FeO accretion is at
first contributed equally by iron oxidized by water flows on the growing
protoplanet and by accretion of FeO-rich planetesimals (that were oxidized by a
similar process on their parent bodies).  Planetesimal accretion nevertheless
dominates after the protoplanet stops accreting ice, but planetesimal accretion
ceases as the protoplanet migrates out of its birth planetesimal ring after 2.3
Myr. The final FeO budget in the mantle is dominated by iron from the
late-accreted pebbles drifting in from the outer Solar System, in agreement with
the match between the iron in CI chondrites and the iron of bulk silicate Earth
\citep{Schiller+etal2020}.

\subsection{Varying the pebble composition}

The final FeO mass fraction depends sensitively on our assumption that the
accreted pebbles were very reduced. This assumption is partially balanced by the
oxidation of this iron by dissolution in hot water; particularly the
early-formed planetesimals deliver a large fraction of the oxidized iron to the
growing protoplanet. While the iron in the chondrules from enstatite chondrites
is nearly 100\% in metallic form, the ordinary chondrites contain approximately
even mixtures of highly reduced Type I chondrules and oxidized Type II
chondrules \citep{Zanda+etal2006}. The chondrules from carbonaceous chondrites
are dominated by reduced Type I chondrules again. This may indicate formation of
chondrules under different oxygen fugacities, with the ordinary chondrite
chondrules tracing a formation region interior of the water ice line where the
water vapour density was high or, alternatively, formation in a denser
environment where the oxygen from sublimated ices would have retained a higher
partial pressure after the flash heating that melted the chondrule precursors,
compared to the carbonaceous chondrite chondrules that formed further from the
Sun \citep{Jacquet+etal2015}.

Our basic assumption of oxidation degree $\zeta_{\rm FeO}=0.05$ and
sulfurization degree $\zeta_{\rm FeS}=0.45$ in the pebbles therefore needs to be
interrogated.  In order to make quick calculations of the final FeO mantle
fraction of planetary bodies under variations of the assumptions, we employ here
a simpler model compared to Figure \ref{f:FeO_t}. We calculate only the {\it
final} FeO fraction based on the simple model: (i) the FeO fraction of the iron
accreted in the first $0.015\,M_{\rm E}$ is the maximum possible given the
sulfur fraction, $\zeta_{\rm FeO}=1-\zeta_{\rm FeS}$, (ii) the FeO fraction of
the accreted planetesimals is the same as the early-accreted material from (i),
and (iii) 50\% of the accreted mass is contributed by planetesimals up to a
planetary mass of $0.15 M_{\rm E}$, the remaining accretion is contributed 100\%
by pebbles. Maps of the FeO mantle fraction are shown in Figure
\ref{f:simple_FeO_model}. We vary $\zeta_{\rm FeO}$ from 0 to 0.2 and
$\zeta_{\rm FeS}$ from 0.2 to 0.5. The FeO mantle fraction of Mars is fit well
in either case; this is due to the large contribution from accreted
planetesimals with high FeO fractions so that the value of $\zeta_{\rm FeO}$
does not affect the final FeO budget of Mars strongly.  Earth, on the other
hand, clearly benefits from a low value of $\zeta_{\rm FeO}=0.05$. The FeO
mantle fraction of Vesta depends strongly on the assumed value of $\zeta_{\rm
FeS}$.  With low values, $\zeta_{\rm FeS}<0.4$, the core mass of Vesta becomes
too small (and the FeO mantle fraction too large).  The terrestrial planets, on
the other hand, have FeO mantle fractions relatively insensitive to the FeS
fraction. As discussed above, planets as massive as Earth and Venus may have
experienced reduced S accretion due to the sublimation of sulfur from FeS in the
hot envelope. Hence, Earth's core may contain only up to one percent of sulfur
\citep{Jackson+etal2021}, while Mars' core could be much more rich in sulfur and
other moderately volatile elements \citep{Stahler+etal2021}.

Overall, our pebble accretion model agrees well with the core mass fractions and
FeO mantle fractions of Vesta, Mars and Earth when the accreted pebbles have a
low degree of oxidation and a high degree of sulfurization. These values are
broadly consistent with constraints from spectra of interstellar dust
\citep{Westphal+etal2019}. The oxidation degree of planetesimals and planets may
therefore be largely a consequence of oxidation of metallic iron on the parent
body. This view in turn implies that the core mass fraction is an increasing
function of the planetary mass, a scaling that will extend also to more massive
super-Earths that form in the inner regions of the protoplanetary disc.

\section{Discussion and implications}
\label{s:discussion}

We demonstrate in this paper how a high primordial ice content of planetesimals
is remembered even after the bulk of the water has been expelled, through the
high fraction of oxidized iron (FeO) in the mantle. \cite{Lichtenberg+etal2019}
proposed that the devolatilization of planetesimals by decay of $^{26}$Al would
lead to a dichotomy of terrestrial planets: planets with a high water fraction
(ocean planets) form in protoplanetary discs with a low initial abundance of
$^{26}$Al, while relatively dry planets akin to Earth emerge when the $^{26}$Al
value is at the solar level or higher. However, accretion of devolatilized
planetesimals would still leave an imprint through their high FeO fractions that
store a significant amount of the oxygen originally present in the ices.  Vesta
has an FeO fraction in the mantle of 25\%--30\%, much higher than the 8\% FeO
mantle fraction of Earth.  Hence it seems unlikely that desiccated planetesimals
akin to Vesta were a major mass source for the accretion of Earth.

Instead, we propose that Vesta belongs to a population of planetesimals that
formed in the earliest stages of the evolution of the solar protoplanetary disc.
Most of these planetesimals were prevented from growing by pebble accretion
after their eccentricities and inclinations were pumped up by close encounters
with one or more of the protoplanets that would later successfully grow to form
Mars, Theia, Earth or even Venus. In some parallel to the model of
\cite{Lichtenberg+etal2019}, we do find that the planetesimal accretion
contribution to these planets could have been dominated by differentiated
objects akin to Vesta. Such planetesimals delivered a significant fraction of
the oxidized FeO to the planets. However, we argue that the bulk delivery of
water to Earth happened through the early accretion of icy pebbles, while the
metallic iron was delivered later with pebbles whose volatile contents
sublimated off while the pebble fell through the envelope of the accreting
planet.

An important implication of the pebble accretion model for terrestrial planet
formation is the emergence of a natural connection between the planetary mass
and the core mass fraction (increasing with mass) as well as the FeO fraction in
the mantle (decreasing with mass). Indeed such a trend is clearly seen for Vesta
(low core fraction, high FeO fraction), Mars (medium core fraction, medium FeO
fraction) and Earth (high core fraction, low FeO fraction). The initial
accretion of highly oxidized material (or rather, material containing a high ice
fraction) followed by later accretion of highly reduced material contrasts with
the prevailing view that the Earth accreted increasingly oxidized material
\citep{WaenkeDreibus1988,WadeWood2005,Rubie+etal2015}. However,
\cite{Badro+etal2015} demonstrated that oxidation models that match the
siderophile element abundance of the terrestrial mantle are not unique and that
solutions can be obtained both for accretion paths that start oxidized and for
those that start reduced. The initially oxidized accretion paths nevertheless
give best fits to the inferred abundance of Si and O in the core in the models
of \cite{Badro+etal2015}. Our pebble accretion model thus agrees well with these
initially oxidized differentiation models.

Our match of the pebble accretion model to the core mass fractions of Vesta,
Mars and Earth implies that the terrestrial planets in the Solar System can be
used to correlate fundamental properties of planets with the planetary mass. Our
model thus predicts that super-Earths formed in the terrestrial planet region
will have very high core mass fractions and low FeO contents in the mantle
(super-Earths forming in the giant planet region could obviously accrete
significant amounts of ice and gases). This may have implications for the
habitability of terrestrial planets and super-Earths, since the core mass
affects the magnetic field and hence the protection of the atmosphere against
the solar wind \citep{Elkins-TantonSeager2008}. The FeO mass fraction, on the
other hand, will determine the flux of Fe$^{2+}$ from the mantle into the
oceans, which affects the oxygen production level needed by the biosphere before
oxygen can build up in the atmosphere rather than being spent on oxidizing the
Fe$^{2+}$ that is continuously dissolved in the oceans. The timing of the great
oxidation of Earth \citep{Lyons+etal2014} may thus have depended on the FeO mass
fraction in the mantle of our planet, which in our view was generously reduced
by the accretion of very dry and metal-rich pebbles.

\begin{acknowledgements}

We thank an anonymous referee for carefully reading the three papers in this
series and for giving us many comments and questions that helped improve the
original manuscripts. We are also grateful to the second referee for
reading the revised manuscript carefully. A.J.\ acknowledges funding from the
European Research Foundation (ERC Consolidator Grant 724687-PLANETESYS), the
Knut and Alice Wallenberg Foundation (Wallenberg Scholar Grant 2019.0442), the
Swedish Research Council (Project Grant 2018-04867), the Danish National
Research Foundation (DNRF Chair Grant DNRF159) and the G\"oran Gustafsson
Foundation.  M.B.\ acknowledges funding from the Carlsberg Foundation
(CF18\_1105) and the European Research Council (ERC Advanced Grant
833275-DEEPTIME). M.S.\ acknowledges funding from Villum Fonden (grant number
\#00025333) and the Carlsberg Foundation (grant number CF20-0209). The
computations were enabled by resources provided by the Swedish National
Infrastructure for Computing (SNIC), partially funded by the Swedish Research
Council through grant agreement no.\ 2020/5-387.

\end{acknowledgements}

\end{document}